\documentclass{article}
\usepackage{spconf,amsmath,graphicx,amssymb}
\usepackage{booktabs}
\usepackage{adjustbox}

\usepackage{multirow}
\usepackage[table,xcdraw]{xcolor}

\usepackage{caption}
\usepackage{subcaption}

\title{Predicting phoneme-level prosody latents using AR and flow-based Prior Networks for expressive speech synthesis }
\name{
\begin{tabular}{c}
	Konstantinos Klapsas$^{\star}$,
	Karolos Nikitaras$^{\star}$,
	Nikolaos Ellinas$^{\star}$, \\
	June Sig Sung$^{\dagger}$,
	Inchul Hwang$^{\dagger}$,
    Spyros Raptis$^{\star}$,
	Aimilios Chalamandaris$^{\star}$,
	Pirros Tsiakoulis$^{\star}$
\end{tabular}
}

\address{$^{\star}$ Innoetics, Samsung Electronics, Greece \\
	$^{\dagger}$ Mobile eXperience Business, Samsung Electronics, Republic of Korea}

\begin{document}
\ninept
\maketitle
\begin{abstract}
A large part of the expressive speech synthesis literature focuses on learning prosodic representations of the speech signal which are then modeled by a prior distribution during inference. In this paper, we compare different prior architectures at the task of predicting phoneme level prosodic representations extracted with an unsupervised FVAE model. We use both subjective and objective metrics to show that normalizing flow based prior networks can result in more expressive speech at the cost of a slight drop in quality. Furthermore, we show that the synthesized speech has higher variability, for a given text, due to the nature of normalizing flows. We also propose a Dynamical VAE model, that can generate higher quality speech although with decreased expressiveness and variability compared to the flow based models.

\end{abstract}

\section{Introduction}
While neural TTS has achieved speech synthesis with very high quality \cite{tacotron,tacotron2} synthesizing expressive speech still remains a difficult challenge. A fundamental difficulty with TTS, and expressive TTS especially, is that the mapping between text and audio is one-to-many, since any sentence can be uttered in a number of different ways. Since neural TTS became a major area of research a lot of attention has been given on expressive speech synthesis by either modeling or controlling the prosody of synthetic speech.

For the purposes of our work we define the prosody of an utterance as any element of it that is not present in the phonetic sequence of the same utterance. Those elements include, but are not limited to $F_0$, Energy and duration. A common approach to this problem is to extract prosodic features from a reference utterance which are then used to condition the decoder of a high quality TTS system. The features can be extracted on either phoneme \cite{Vioni2021ProsodicCF, Lee2019RobustAF}, word \cite{Klapsas2021WordLevelSC, Guo2022UnsupervisedWP} or sentence level \cite{pitch_loud, Wang2018StyleTU}.

A very influential work on this field is FVAE \cite{fully_hierarchical} which extracts fine grained prosodic features utilizing a VAE framework. The drawback of this model is that since the prior distribution it assumes for the latent variables is independent for each phoneme, it is impossible to use this model for TTS and generate high quality samples, since in real speech, the prosodic characteristics of the phonemes are not independent of each other and any synthesized speech with this property will sound unnatural. Thus, the only use for models like this is when there is reference utterance at inference time, or for modifying the prosody of an given utterance.

This problem is general for any approach more fine-grained than utterance level, and as a solution, a lot of work utilises extra modules that predict the prosodic features, given the text and possibly some additional features \cite{Klapsas2021WordLevelSC, sun2020generating, proso, Chien2021HierarchicalPM, Stanton2018PredictingES}. 

A framework which has recently gained traction in TTS is the normalizing flows framework \cite{Kingma2018GlowGF}. It has been used to replace the decoder in a TTS system \cite{Kim2020GlowTTSAG} and applied to a prior distribution of a VAE model in order to increase its expressiveness in an end-to-end TTS system \cite{vits}. It has also been applied to the problem of duration modeling, which is an important subset of the task of expressive speech synthesis \cite{expresrive} . Normalizing flows are promising since they are invertible models with tractable likelihoods, and thus you can train them with the exact maximum likelihood. They also allow for efficient sampling of the base distribution, something that is of particular use in the one-to-many problem of expressive TTS.

In this work, we use the posterior latents of an FVAE system as unsupervised prosodic features that we wish to predict from the text in order to be able to synthesize without needing a reference. We investigate the performance of different architectures at that task, namely Autoregressive (AR) models and flow based models. We also modify the FVAE architecture to have a prior distribution that is both autoregressive and dependent on the text and thus, instead of being independent for each phoneme, captures the dependencies of the prosody of natural speech. We evaluate the models both with MOS and with a number of other metrics that attempt to measure expressiveness of a TTS system and how diverse are the samples it generates for the same text.

\section{Method}

% \begin{minipage}[b]{.5\textwidth}  %1.0\linewidth}
%   \centering
%   \centerline{\includegraphics[width=.4\linewidth]{vae.png}}     %[width=8.5cm]{vae.png}}
% %  \vspace{2.0cm}
%   \centerline{(a) Result 1}  %\medskip
% \end{minipage}
% %
% \begin{minipage}[b]{.5\textwidth}         %{.48\linewidth}
%   \centering
%   \centerline{\includegraphics[width=.4\linewidth]{flow.png}}       %width=4.0cm]{flow.png}}
% %  \vspace{1.5cm}
%   \centerline{(b) Results 3}\medskip
% \end{minipage}
% \hfill

% \begin{figure}[ht]

% \centering
% \begin{minipage}{\textwidth}
%   \centering
%   \includegraphics[width=.4\linewidth]{vae.png}
%   \captionof{figure}{A figure}
%   \label{fig:test1}
% \end{minipage}%
% \begin{minipage}{\textwidth}
%   \centering
%   \includegraphics[width=.4\linewidth]{flow.png}
%   \captionof{figure}{Another figure}
%   \label{fig:test2}
% \end{minipage}
% \end{figure}

% %
% \caption{Example of placing a figure with experimental results.}
% \label{fig:res}
% %
% \end{figure}

% \begin{figure}
% \centering
% \begin{subfigure}{0.5\textwidth}
%     \includegraphics[width=\textwidth]{vae.png}
%     \caption{First subfigure.}
%     \label{fig:first}
% \end{subfigure}
% \hfill
% \begin{subfigure}{0.5\textwidth}
%     \includegraphics[width=\textwidth]{flow.png}
%     \caption{Second subfigure.}
%     \label{fig:second}
% \end{subfigure}
        
% \caption{Creating subfigures in \LaTeX.}
% \label{fig:figures}
% \end{figure}

\begin{figure*}[!ht]
\centering
\includegraphics[width=\textwidth]{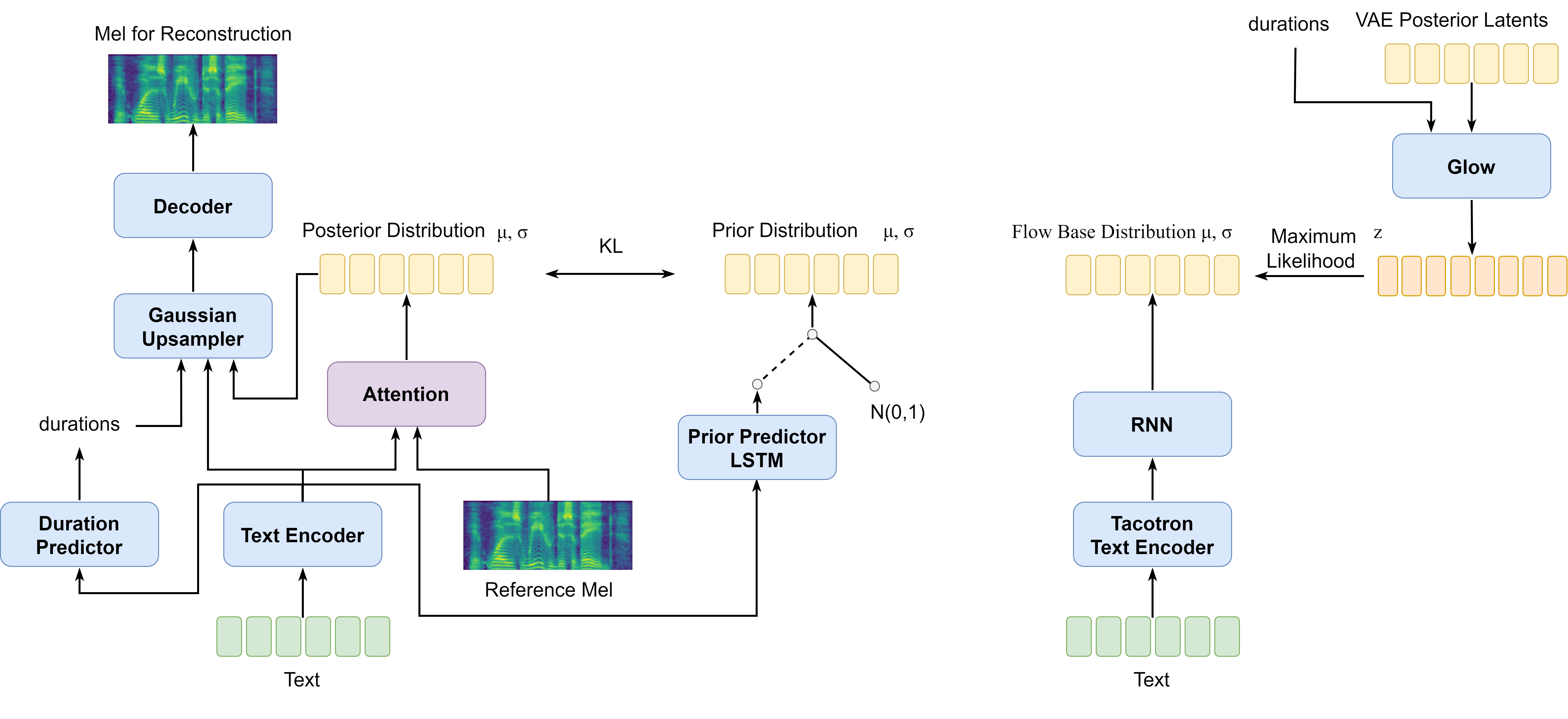}
\caption{Architecture of the FVAE model, and the two prior distributions, N(0,I) and the AR prior distribution for the DVAE model. On the right we see an overview of the flow model used to maximise the likelihood of the posterior latents. The AR predictor that predicts the samples from N(0,I) is not pictured.}
\label{fig:full}
\end{figure*}

\subsection{Fine Grained VAE}

We use a Fine Grained VAE \cite{Lee2019RobustAF, fully_hierarchical} to get latent prosodic representations for our speech signals. It is based on a Non-Attentive Tacotron model \cite{non_attentive} with an extra phoneme level encoder that uses a simple Location Sensitive Attention \cite{attention} to align the phoneme sequence from the Tacotron Encoder output with the reference mel spectrogram. This representation parameterizes the phoneme level posterior distribution that hopefully captures the prosodic information of the utterance. 

The goal of the model is to maximize the evidence lower bound (ELBO) which is formulated as follows:

\begin{equation}\label{elbo}
    \mathcal{L} = \mathbb{E}_{q(\mathbf{z}|\mathbf{x})} \log p(\mathbf{x}|\mathbf{z}) - \sum_{n=1}^{N} D_{KL} \big(q(\mathbf{z_n} | \mathbf{x}, \mathbf{y}) || p(\mathbf{z_n})\big)
\end{equation}
where $\mathbf{z_n}$ is the latent distribution of the nth phoneme,$\mathbf{x}, \mathbf{y}$ are the mel-spectrogram and the text respectively, and q and p are the posterior and prior distributions. %It should be noted that in this formulation the prior distribution is allowed to be dependent on the text.

\subsection{Autoregressive Prior}
The most common choice for the prior distribution is the standard Gaussian $\mathcal{N}(0,I)$. This distribution however cannot be used at inference since it implicitly assumes independence for the latents of each phoneme, something that is not true of the posterior distribution. As a result speech where the latents are sampled from this distribution is quite unnatural and an extra AR prior network is used to capture the temporal aspect of the posterior distribution \cite{sun2020generating}. In our work we consider two architectures for implementing the prior autoregressive network.

The first approach is to simply use $\mathcal{N}(0,I)$ during the training of the VAE model and then train an extra model that given the text, predicts a Gaussian distribution whose KL divergence from the posterior distribution is then minimized. The model we use is an LSTM module which uses the posterior latents as a teacher forced input. 

The second approach is to directly modify the prior in eq. \ref{elbo} to an autoregressive distribution, thereby making the model a Dynamical Variational Autoencoder (DVAE) \cite{vrnn, storn, dvae}. In general, the modification for the KL term is as follows:

\begin{equation}\label{kl} 
    \mathcal{L}_{KL} = \sum_{n=1}^{N} D_{KL} \big(q(\mathbf{z_n} | \mathbf{z_{<n}}, \mathbf{x}, \mathbf{y}) || p(\mathbf{z_n}| \mathbf{z_{<n}},\mathbf{y})\big) 
\end{equation}

In practice, since we find that the posterior latents are capable of reconstructing the signal quite well in the baseline model, we do not change the reference encoder, and therefore do not make use of the autoregressive connections implied by the loss in eq. \ref{kl}, that is to say we assume $q(\mathbf{z_n} | \mathbf{z_{<n}}, \mathbf{x}, \mathbf{y}) = q(\mathbf{z_n} | \mathbf{x}, \mathbf{y}) $. The prior distribution is changed to an AR model with the same architecture as the one in the previous approach but this time trained concurrently with the rest of the VAE. We do find however that fine-tuning only the prior predictor after the rest of the model has finished training, (keeping all the other modules fixed) leads to significantly better performances.

We parameterise both the posterior and the prior distributions as isotropic Gaussian distributions, which gives the KL divergence the form, for each latent dimension: 
\begin{equation}\label{kl_gauss}
    \mathcal{L}_{KL} = \log{\frac{\sigma_p}{\sigma_q}} + \frac{\sigma_q^2 + (\mu_p - \mu_q)^2}{2\sigma_p^2}  - \frac{1}{2}
\end{equation}
where $q(\mathbf{z_n} | \mathbf{x}, \mathbf{y})  = \mathcal{N}(\mu_q(\mathbf{x}, \mathbf{y}),\sigma_q(\mathbf{x}, \mathbf{y})) $ and $p(\mathbf{z_n} | \mathbf{z_{<n}}, \mathbf{y})  = \mathcal{N}(\mu_p (\mathbf{z_{<n}}, \mathbf{y}),\sigma_q(\mathbf{z_{<n}}, \mathbf{y})) $.

\subsection{Normalizing Flows }
A different way to predict the posterior latents is by training a normalizing flow model to maximise the likelihood of the latents sampled from the VAE posterior distribution. We first train a Fine Grained VAE model with a $\mathcal{N}(0,I)$ as prior and then we save all of the posterior latents of all the utterances in the dataset, in order to train our flow model. The base distribution of the flow model is an isotropic Gaussian distribution with parameters that depend on the phonetic information. During inference, we simply sample from this distribution and get a corresponding sequence of latents, which are then used on the decoder of the pre-trained VAE model.

Since phoneme durations are also part of the prosody, we concatenate the posterior latents with the ground truth durations for each utterance, as inputs to the flow model, during training. Therefore, our flow models also maximise the likelihood of the duration sequence, and consequently perform duration prediction in a similar way as in \cite{expresrive}.

A normalizing flow consists of a series of invertible transformations $f_i, i = 0 \ldots M$, which are composed to transform from the base distribution to the desired distribution as:
\begin{equation}
\mathbf{z} = f_{M} \circ \ldots f_{1} \circ f_{0} (\mathbf{z_0}), \quad \mathbf{z_0}  \sim p_0(\mathbf{z_0}| \mathbf{c})
\end{equation}
where $\mathbf{z}$ is the posterior of the pre-trained VAE, $p_0$ is the base distribution and $\mathbf{c}$ is the conditioning phonetic information. In our experiments we use as $c$ the phoneme encoder outputs from the pre-trained VAE model. We pass those outputs through an RNN model and with two linear layers we parameterise the mean and standard deviation of the base distribution. The exact log-likelihood of the normalizing flow model is tractable and can be directly optimized with gradient descent. It is given by the equation:

\begin{equation}\label{flow}
   \log{p_M(\mathbf{z}| \mathbf{c}) } = \log{p_0(\mathbf{z_0}| \mathbf{c})} + \sum_{i=0}^M \log{| \det(\mathbf{J} (f_i(\mathbf{z_{i}} ))) |^{-1}}
\end{equation}
where  $ | \det (\mathbf{J}) |$ is the absolute value of the determinant of the Jacobian matrix. 

We conjecture that the flow models will help with learning more variable speech since they are less prone to collapsing than AR models which are based on Kullback Leibler Divergence optimization.

\section{Evaluation}

\subsection{Experiment settings}
We train and test our models on a subset of the Blizzard 2013 Challenge single-speaker audiobook dataset \cite{King2014TheBC} which contains 85 hours of speech with highly varying prosody. We base our TTS model on a Non-attentive \cite{non_attentive} variant of \cite{lpctron} with an LPCNet vocoder \cite{lpcnet} and we use the VAE reference encoder from \cite{fully_hierarchical} but with the difference that the latent dimensions are sampled at once instead of hierarchically. Our flow models are based on Glow \cite{Kingma2018GlowGF, Kim2020GlowTTSAG}.

Our VAE models are trained for 350k iterations, while the training of the prior models, both for the AR models and the fine-tuning of the dynamical models is being done for 100k iterations. Audio samples from all our models are available at our demo page \footnote{https://innoetics.github.io/publications/phoneme-flow-prior/index.html}.

\subsection{Evaluation} 

We compare our models with the following baselines: 
\begin{itemize}
  \item A simple Non-attentive model with no prosody modeling (NAT)
  \item VITS \cite{vits}
  \item A Non-attentive model where the duration predictor module has been replaced with the stochastic duration predictor from VITS
\end{itemize}

 The stochastic duration prediction module is a flow based model that utilizes variational data augmentation \cite{variational_augm} to enable high dimensional transformation in normalizing flows models, and variational dequantization \cite{dequant} to deal with the discrete nature of duration data.

\subsubsection{Reconstruction}
We compare the reconstruction performances of the VAE models, using the metrics $F_0$ frame error (FFE) \cite{FFE}, and mel-cepstral distortion (MCD) \cite{MCD} for different dimensions of the latent variables. The models compared are the Dynamical VAE models and the baseline FVAE model, and the reconstruction is being done with the posterior latents.

\begin{table}[th]
\caption{Reconstruction. Lower is better.}
\label{tab:recon}
\centering
\begin{tabular}{cccc}
\toprule
                         &                                        &                              &                               \\
latent dim          & model                                  & MCD                          & FFE                           \\ \midrule

                         & FVAE                            & 3.85                         & 5.97                          \\
\multirow{-2}{*}{8-dim}  & DVAE                        & 3.88                         & 6.24                          \\ \midrule
                         & FVAE                            & 3.31                         & 4.64                          \\
\multirow{-2}{*}{16-dim} & DVAE                        & 3.30                         & 4.83                          \\ \midrule
                         & FVAE                            & 3.02                         & 4.44                          \\ 
\multirow{-2}{*}{32-dim} & DVAE                        & 3.03                         & 4.44                                           \\  \midrule
                         & FVAE                            & 2.93                         & 4.45                          \\
\multirow{-2}{*}{64-dim} & DVAE                        & 2.89                         & 4.40                          \\
\bottomrule
\end{tabular}
\end{table}

The results can be seen in Table \ref{tab:recon}. We observe that 1) as expected, increasing the dimensions of the latent space improves the reconstruction performance of the models 2) the improvement is marginal for large dimensions and 3) whether the model was trained on the $\mathcal{N}(0,I)$ prior or the prior predicted from the text, makes little difference. 

\subsubsection{Naturalness and Expressiveness}\label{expr_nat_section}

Following \cite{sun2020generating} we evaluate the expressiveness of the utterances by reporting the standard deviations of three phoneme level prosody features: Energy, $F_0$ and duration. We first align each sentence with the phonemes of the corresponding text using Kaldi \cite{kaldi} and then we extract the three features for each phoneme, by using the number of frames in the case of the duration, and by averaging the features across the frames that correspond to the phoneme in the other two cases. $F_0$ extraction was done with the Praat toolkit \cite{praat}. We then group the phonemes across all of the validation utterances,  calculate the standard deviations of the three features and then average them for the different phonemes. 

% \begin{equation}
%     Prosody_std(i) = 
% \end{equation}
We also conduct a MOS naturalness test to evaluate the performance of our models, by crowd-sourcing native English listeners to rate the utterances on a 1-5 scale. To improve the quality of the evaluations, besides using validation utterances, we excluded all pages where the ground truth utterance was given score less than 3 or pages where the mean synthetic score was higher than that of the ground truth.

We also report the Word Error Rate (WER) of a pre trained ASR model \cite{Shi2021EmformerEM}. While the WER is a measure of how well the sentence contains the content of the text, we expect that sentences with unnatural prosody will tend to have higher WER, and as such we can use it as an additional metric of naturalness.

We experiment with two dimensions for our models, 8 and 16 and we also investigate the effect of sampling from the base distribution using a smaller standard deviation to all the metrics. In particular, we use 0.33, 0.5 and 0.8 as stds for the base Gaussian.

All of the metrics, both subjective and objective, were evaluated on a unseen during training (by both the FVAE and the prior architectures) subset with 50 sentences.

\begin{table}[th]
\caption{Naturalness and Expressiveness Metrics.}
\label{tab:nat_exp}
\centering
\begin{adjustbox}{width=0.5\textwidth}
\begin{tabular}{c|c|c|ccc}
\toprule
  model            & MOS      &   WER   & \multicolumn{3}{c}{Prosody stddev} \\

             & &   & E         & F0         & Dur       \\
\midrule
Real Speech   &   $4.51 \pm 0.06  $ & 4.5    & 5.25        & 44.62         &    43.21       \\
NAT           &   $3.30 \pm 0.11 $  & 5.8  & 3.49      & 18.26      & 18.54     \\
NAT w/ flow dur\_pred  & $3.19 \pm 0.11$ & 5.3  & 4.20      & 27.08      & 39.67     \\
VITS        &      $3.16 \pm 0.11$   & 6.7  & 4.31      & 28.86      & 29.01     \\ \midrule
FVAE  8    & $ 3.03 \pm 0.11$      & 6.1  & 3.89      & 23.77      & 20.13     \\
DVAE 8     &  $ 3.26 \pm 0.11 $  & 5.4  & 3.76      & 21.38      & 19.64     \\ \midrule
Flow 8 dim std 0.33       &  $2.98 \pm 0.11$  & 6.1  & 3.76      & 25.62      & 21.39     \\
Flow 8 dim std 0.5     &  $ 2.82 \pm 0.11$  & 6.3  & 4.04      & 28.94      & 23.09     \\
Flow 8 dim std 0.8     &  $ 2.53 \pm 0.11$  & 7.1  & 4.55      & 35.59      & 28.75     \\\midrule
FVAE 16     &  $ 2.99 \pm 0.11$ & 4.9  & 3.88      & 20.51      & 19.31     \\
DVAE 16    &  $ 3.19 \pm 0.11$  & 5.5  & 3.76      & 24.26      & 19.56     \\ \midrule
Flow 16 dim std 0.33        &  $ 2.85 \pm 0.11$  & 5.3  & 3.75      & 22.93      & 20.32     \\
Flow 16 dim std 0.5         &  $ 2.88 \pm 0.12$  & 6.2  & 4.02      & 26.68      & 22.09     \\
Flow 16 dim std 0.8         &  $2.41\pm 0.11 $  & 8.3  & 4.51      & 34.34      & 26.50     \\
  \bottomrule
\end{tabular}
\end{adjustbox}
\end{table}

We can see the results in Table \ref{tab:nat_exp}. 
The system with the best naturalness score is the simple NAT. However it has the lowest score on all of the prosody metrics, especially compared to the real speech samples which have the highest score on the same metrics. We conclude that NAT can synthesize very natural sounding samples but with ``flat" prosody.

We suspect that the relatively low numbers of VITS are due to the fact that the dataset is highly expressive and thus substantially more difficult to model, compared to most datasets. We do note however, that we did not experiment with fine-tuning the standard deviations of the flow models for VITS, for this dataset. 

In general, we observe that there is a big trade-off between naturalness and expressiveness. We can see this directly by comparing the samples of the flow models with different standard deviations. We notice that the larger standard deviation is used for the base distribution, the larger prosody standard deviation we get in the output, but with a corresponding deterioration of the MOS score and the WER. 

Another point worth noting, is that while the extra dimensions increase the reconstruction performance of the FVAE, for both approaches of predicting the latents, both the MOS score and the prosody std worsen. We can attribute that on the high dimensional space of the latents being harder to predict. The conclusion of this observation is that the bottleneck of this approach is on the prior network rather than on the FVAE, especially given how good the posterior reconstructions are.

The performance of the NAT model with the stochastic duration predictor is perhaps surprising, since we not only see a large variation in the phoneme durations, but also in the other prosodic features, despite not performing any explicit modeling on them or any other unsupervised prosody features. A possible explanation of this, is that the phoneme durations are highly correlated with the other features and that by sampling from a more rich prior duration distribution we encourage the decoder to create variations in the correlated features as well. 

Excluding NAT, the best MOS results both come from the DVAE models, even though they are both within the confidence intervals for VITS and NAT with flow duration predictor. In particular, the DVAE models have significantly higher MOS scores than the FVAE baselines, without a significant compromise in the rest of the metrics. We should note that beside the MOS scores, the DVAE models outperform NAT in all other metrics.

Finally, by comparing the flow based networks with the two AR networks, we see again the same trade-off, namely that while flow networks have better results in the prosodic features metrics, we see both lower MOS and higher WER even for low std values in the base distribution.
  
\subsubsection{Diversity}
Since the mapping from text to speech is one-to-many, an ideal model should generate utterances with different prosody for the same given text. While, in general, the prosody of a sentence depends on semantic properties of the text, and thus cannot be said to be independent of it, we expect that for models with similar MOS naturalness scores, prosodic variations are actually desirable. That is, we expect that a failure of any system to generate natural prosody for a given sentence will be reflected in the MOS score of that system.

In order to evaluate the diversity of the generated utterances, we simply sample from the prior model a number of times, with the same text as input, and then calculate the three prosodic features of each phoneme by aligning the phonemes with each sampled utterance, in the same manner as in section \ref{expr_nat_section}. This time, we do not group the phonemes by type, but by position in the utterance. We then find the standard deviation for each feature, across all of the different synthesized versions of each utterance, and then finally average them over all the phonemes and all the different texts. We use 50 sentences, each sampled 10 times from every model.

\begin{table}[th]
\caption{Diversity Metrics.}
\label{tab:diversity}
\centering
\begin{tabular}{c|ccc}
\toprule
                    & \multicolumn{3}{c}{Prosody stddev} \\
model               & E        & F0       & Dur      \\
\midrule
NAT + flow dur\_pred & 2.17     & 16.31    & 16.21    \\
VITS                & 1.87     & 14.78    & 9.59     \\\midrule
FVAE  dim 8      &     0.67     &   6.90       &  3.59        \\
DVAE  dim 8   &    0.63      &   6.01       &   3.06       \\\midrule
Flow 8 dim std 0.33         & 1.52     & 13.40    & 6.07     \\
Flow 8 dim std 0.5          & 2.09     & 17.70    & 8.02     \\
Flow 8 dim std 0.8          & 3.06     & 26.48    & 12.03    \\\midrule
FVAE  dim 16       &    0.61      &   5.60       &    2.68      \\
DVAE  dim 16   &     0.60     &   5.20       &    2.69      \\\midrule
Flow 8 dim std 0.33         & 1.52     & 12.40    & 5.68     \\
Flow 8 dim std 0.5          & 2.06     & 17.07    & 7.42     \\
Flow 8 dim std 0.8          & 3.02     & 25.84    & 11.28    \\
\bottomrule
\end{tabular}
\end{table}

We can see that the diversity of the AR predictors, both for the FVAE and the DVAE cases is quite poor, presumably because of some form of posterior collapse in the training of the VAE models, regardless of the prior. That is to say, in order for the model to improve the reconstruction loss, it decreases the standard deviation of the posterior distribution, which in turn corresponds to a lower std after fine-tuning. We do not notice this problem in the flow based models, because since the transformation is guaranteed to be invertible, there isn't an opportunity for the model to collapse different samples from the base distribution to the same output. This observation also holds for model with similar MOS scores (e.g. FVAE 8 and Flow 8 std 0.33).

\section{Conclusions}
We showed different architectures for predicting prosody latents in an FVAE model. We demonstrated that using flow based networks can generate prosodically diverse utterances but with somewhat decreased naturalness, while using Dynamical VAE models or even a simple AR prior, will give to higher quality results but with less variability. So far, the trade-off between greater expressiveness and variability and more natural utterances seems to be an important obstacle to approaching the characteristics of real speech. Since we show evidence that the bottleneck is on the prior prediction, future work should improve on the prior networks, flow or otherwise.

\clearpage
%\subsection{}  

\bibliographystyle{IEEEbib}
\bibliography{strings,refs}

\begin{thebibliography}{10}

\bibitem{tacotron}
Yuxuan~Wang \textit{et al.},
\newblock ``{Tacotron: Towards End-to-End Speech Synthesis},''
\newblock in {\em Proc. Interspeech}, 2017.

\bibitem{tacotron2}
Jonathan~Shen \textit{et al.},
\newblock ``Natural tts synthesis by conditioning wavenet on mel spectrogram
  predictions,''
\newblock in {\em Proc. ICASSP}, 2018.

\bibitem{Vioni2021ProsodicCF}
Alexandra Vioni, Myrsini Christidou, Nikolaos Ellinas, Georgios Vamvoukakis,
  Panos Kakoulidis, Taehoon Kim, June~Sig Sung, Hyoungmin Park, Aimilios
  Chalamandaris, and Pirros Tsiakoulis,
\newblock ``{Prosodic Clustering for Phoneme-Level Prosody Control in
  End-to-End Speech Synthesis},''
\newblock 2021.

\bibitem{Lee2019RobustAF}
Younggun Lee and Taesu Kim,
\newblock ``{Robust and Fine-grained Prosody Control of End-to-end Speech
  Synthesis},''
\newblock in {\em Proc. ICASSP}, 2019.

\bibitem{Klapsas2021WordLevelSC}
Konstantinos Klapsas, Nikolaos Ellinas, June~Sig Sung, Hyoungmin Park, and
  Spyros~N. Raptis,
\newblock ``Word-level style control for expressive, non-attentive speech
  synthesis,''
\newblock in {\em Proc. SPECOM}, 2021.

\bibitem{Guo2022UnsupervisedWP}
``{Unsupervised Word-Level Prosody Tagging for Controllable Speech
  Synthesis},''
\newblock in {\em Yiwei Guo and Chenpeng Du and Kai Yu}, 2022.

\bibitem{pitch_loud}
Siddharth Gururani, Kilol Gupta, Dhaval Shah, Zahra Shakeri, and Jervis Pinto,
\newblock ``Prosody transfer in neural text to speech using global pitch and
  loudness features,''
\newblock 11 2019.

\bibitem{Wang2018StyleTU}
Yuxuan Wang, Daisy Stanton, Yu~Zhang, R.~Skerry-Ryan, Eric Battenberg, Joel
  Shor, Y.~Xiao, F.~Ren, Ye~Jia, and R.~A. Saurous,
\newblock ``Style tokens: Unsupervised style modeling, control and transfer in
  end-to-end speech synthesis,''
\newblock in {\em Proc. ICML}, 2018.

\bibitem{fully_hierarchical}
Guangzhi Sun, Heiga Zen, Ron~J. Weiss, Yonghui Wu, Yu~Zhang, and Yuan Cao,
\newblock ``Fully-hierarchical fine-grained prosody modeling for interpretable
  speech synthesis,''
\newblock in {\em Proc. ICASSP}, 2020.

\bibitem{sun2020generating}
Guangzhi Sun, Y.~Zhang, Ron~J. Weiss, Yuan Cao, H.~Zen, A.~Rosenberg,
  B.~Ramabhadran, and Yonghui Wu,
\newblock ``Generating diverse and natural text-to-speech samples using a
  quantized fine-grained vae and autoregressive prosody prior,''
\newblock in {\em Proc. ICASSP}, 2020.

\bibitem{proso}
Yi~Ren, Ming Lei, Zhiying Huang, Shiliang Zhang, Qian Chen, Zhijie Yan, and
  Zhou Zhao,
\newblock ``{Prosospeech: Enhancing Prosody with Quantized Vector Pre-Training
  in Text-To-Speech},''
\newblock in {\em Proc. ICASSP}, 2022.

\bibitem{Chien2021HierarchicalPM}
C.~M. Chien and Hung yi~Lee,
\newblock ``{Hierarchical Prosody Modeling for Non-Autoregressive Speech
  Synthesis},''
\newblock in {\em Proc. SLT}, 2021.

\bibitem{Stanton2018PredictingES}
Daisy Stanton, Yuxuan Wang, and R.~Skerry-Ryan,
\newblock ``Predicting expressive speaking style from text in end-to-end speech
  synthesis,''
\newblock in {\em Proc. SLT}, 2018.

\bibitem{Kingma2018GlowGF}
Diederik~P. Kingma and Prafulla Dhariwal,
\newblock ``Glow: Generative flow with invertible 1x1 convolutions,''
\newblock {\em ArXiv}, vol. abs/1807.03039, 2018.

\bibitem{Kim2020GlowTTSAG}
Jaehyeon Kim, Sungwon Kim, Jungil Kong, and Sungroh Yoon,
\newblock ``Glow-tts: A generative flow for text-to-speech via monotonic
  alignment search,''
\newblock {\em ArXiv}, vol. abs/2005.11129, 2020.

\bibitem{vits}
Jaehyeon Kim, Jungil Kong, and Juhee Son,
\newblock ``Conditional variational autoencoder with adversarial learning for
  end-to-end text-to-speech,''
\newblock in {\em Proc. ICML}, 2021.

\bibitem{expresrive}
Syed Abbas, Thomas Merritt, Alexis Moinet, Sri Karlapati, Ewa Muszynska, Simon
  Slangen, Elia Gatti, and Thomas Drugman,
\newblock ``{Expressive, Variable, and Controllable Duration Modelling in
  TTS},''
\newblock in {\em Proc. Interspeech}, 2022.

\bibitem{non_attentive}
Jonathan Shen, Ye~Jia, M.~Chrzanowski, Yanshun Zhang, I.~Elias, H.~Zen, and
  Yonghui Wu,
\newblock ``Non-attentive tacotron: Robust and controllable neural tts
  synthesis including unsupervised duration modeling,''
\newblock {\em ArXiv}, vol. abs/2010.04301, 2020.

\bibitem{attention}
Ashish Vaswani, Noam Shazeer, Niki Parmar, Jakob Uszkoreit, Llion Jones,
  Aidan~N Gomez, \L~ukasz Kaiser, and Illia Polosukhin,
\newblock ``{Attention is All you Need},''
\newblock in {\em Advances in Neural Information Processing Systems}, 2017.

\bibitem{vrnn}
Junyoung Chung, Kyle Kastner, Laurent Dinh, Kratarth Goel, Aaron~C Courville,
  and Yoshua Bengio,
\newblock ``A recurrent latent variable model for sequential data,''
\newblock in {\em Advances in Neural Information Processing Systems},
  C.~Cortes, N.~Lawrence, D.~Lee, M.~Sugiyama, and R.~Garnett, Eds., 2015.

\bibitem{storn}
Justin Bayer and Christian Osendorfer,
\newblock ``Learning stochastic recurrent networks,''
\newblock {\em ArXiv}, vol. abs/1411.7610, 2014.

\bibitem{dvae}
Laurent Girin, Simon Leglaive, Xiaoyu Bie, Julien Diard, Thomas Hueber, and
  Xavier Alameda-Pineda,
\newblock ``Dynamical variational autoencoders: A comprehensive review,''
\newblock {\em Foundations and Trends® in Machine Learning}, vol. 15, no. 1-2,
  pp. 1--175, 2021.

\bibitem{King2014TheBC}
S.~King and Vasilis Karaiskos,
\newblock ``The blizzard challenge 2013,''
\newblock in {\em Blizzard Challenge Workshop}, 2013.

\bibitem{lpctron}
Nikolaos~Ellinas \textit{et al},
\newblock ``High quality streaming speech synthesis with low,
  sentence-length-independent latency,''
\newblock in {\em Proc. Interspeech}, 2020.

\bibitem{lpcnet}
J.~{Valin} and J.~{Skoglund},
\newblock ``{LPCNet: Improving Neural Speech Synthesis through Linear
  Prediction},''
\newblock in {\em Proc. ICASSP}, 2019.

\bibitem{variational_augm}
Jianfei Chen, Cheng Lu, Biqi Chenli, Jun Zhu, and Tian Tian,
\newblock ``{VF}low: More expressive generative flows with variational data
  augmentation,''
\newblock in {\em Proceedings of the 37th International Conference on Machine
  Learning}, Hal~Daumé III and Aarti Singh, Eds. 13--18 Jul 2020, vol. 119 of
  {\em Proceedings of Machine Learning Research}, pp. 1660--1669, PMLR.

\bibitem{dequant}
Jonathan Ho, Xi~Chen, Aravind Srinivas, Yan Duan, and Pieter Abbeel,
\newblock ``Flow++: Improving flow-based generative models with variational
  dequantization and architecture design,''
\newblock in {\em Proc. ICML}, 2019.

\bibitem{FFE}
W.~Chu and A.~Alwan,
\newblock ``{Reducing F0 Frame Error of F0 tracking algorithms under noisy
  conditions with an unvoiced/voiced classification frontend},''
\newblock in {\em Proc. ICASSP}, 2009.

\bibitem{MCD}
R.~Kubichek,
\newblock ``Mel-cepstral distance measure for objective speech quality
  assessment,''
\newblock {\em Proceedings of IEEE Pacific Rim Conference on Communications
  Computers and Signal Processing}, vol. 1, pp. 125--128 vol.1, 1993.

\bibitem{kaldi}
\textit{et al.} Povey,
\newblock ``The kaldi speech recognition toolkit,''
\newblock in {\em Proc. ASRU}, 2011.

\bibitem{praat}
R.~Corretge,
\newblock ``Praat vocal toolkit,'' 2012-2020.

\bibitem{Shi2021EmformerEM}
Yangyang Shi, Yongqiang Wang, Chunyang Wu, Ching feng Yeh, Julian Chan, Frank
  Zhang, Duc Le, and Michael~L. Seltzer,
\newblock ``Emformer: Efficient memory transformer based acoustic model for low
  latency streaming speech recognition,''
\newblock {\em Proc. ICASSP}, 2021.

\end{thebibliography}

\end{document}